\documentclass[a4paper,11pt,titlepage]{article}
\usepackage{graphicx}
\usepackage{epsfig}
\usepackage{amssymb,amsfonts} 
\begin{document}

\begin{titlepage}

\begin{flushright}
UTAS-PHYS-2008-09\\
May 2008\\
\end{flushright}

\begin{centering}
 
{\ }\vspace{0.5cm}

\LARGE\textbf{Regularised Tripartite Continuous Variable EPR-type States with Wigner Functions and CHSH Violations}

\vspace{0.5cm}
\normalsize\textbf{Sol H. Jacobsen\footnote{Commonwealth Endeavour Scholar} and P.D. Jarvis\footnote{Alexander von Humboldt Fellow}}

\vspace{0.3cm}

{\em School of Mathematics and Physics}\\
{\em University of Tasmania, Private Bag 37}\\
{\em 7001 Hobart, Tasmania, Australia}\\
{\em E-mail: {\tt solj@utas.edu.au}, {\tt Peter.Jarvis@utas.edu.au}}

\normalsize 
\end{centering}

\section*{Abstract}
We consider tripartite entangled states for continuous variable systems of EPR type, which generalise the famous bipartite CV EPR states (eigenvectors of conjugate choices $X_1 - X_2, P_1+ P_2$, of the systems' relative position and total momentum variables). We give the regularised forms of such tripartite EPR states in second-quantised formulation, and derive their Wigner functions. This is directly compared with the established NOPA-like states from quantum optics. Whereas the multipartite entangled states of NOPA type have singular Wigner functions in the limit of large squeezing, $r \rightarrow \infty$, or $\tanh  r \rightarrow 1^-$ (approaching the EPR states in the bipartite case), our regularised tripartite EPR states show singular behaviour not only in the approach to the EPR-type region ($s \rightarrow 1$ in our notation), but also for an additional, auxiliary regime of the regulator ($s \rightarrow \sqrt{2}$). While the $s\rightarrow 1$ limit pertains to tripartite CV states with singular eigenstates of the relative coordinates and remaining squeezed in the total momentum, the $s\rightarrow \sqrt{2}$ limit yields singular eigenstates of the total momentum, but squeezed in the relative coordinates. Regarded as expectation values of displaced parity measurements, the tripartite Wigner functions provide the ingredients for generalised CHSH inequalities. Violations of the tripartite CHSH bound ($B_3 \le 2$) are established, with $B_3 \cong 2.09$ in the canonical regime ($s \rightarrow 1^+$), as well as $B_3 \cong 2.32$ in the auxiliary regime ($s \rightarrow \sqrt{2^+}$).


\end{titlepage}

\section{Introduction}

The nature of quantum entanglement has been pursued almost since the inception of quantum mechanics itself. Whereas the early insights of Schr\"{o}dinger, as well as of Einstein, Podolsky and Rosen (EPR) \cite{EPR1935} were framed in the \emph{Gedankenexperiment} mode of discussion with continuous degrees of freedom (particle position and momentum eigenstates), the issues were taken up quantitatively in Bell's theorem \cite{Bell1964} for the case of spin degrees of freedom, via the transcription to this context given by Bohm \cite{Bohm1951}. However, continuous variable (CV) systems are the natural framework for most quantum optics and quantum communication work \cite{NielsenandChuang1968, WallsandMilburn1994}, and Bell, and the more general Clauser, Horne, Shimony and Holt (CHSH) inequalities \cite{Clauseretal1969}, are important measures of entanglement. A technical difficulty in working with continuous variables is that the theoretical ideal EPR type states are singular, whereas experimental investigations require regularised states. In the bipartite case these may be provided by so-called NOPA squeezed states \cite{Reid1989, Ouetal1988, Ouetal1992, BanaszekandWodkiewicz1998, BanaszekandWodkiewicz1999} in the large squeezing limit. In an alternative approach, Fan and Klauder \cite{FanandKlauder1994} constructed somewhat more general classes of EPR states, but without providing a regularisation.

In the extension to multipartite cases, a natural question is the choice of relative variables, chosen from amongst the positions and momenta of the constituent particles of the system, which will provide an appropriate generalisation of the bipartite EPR states (with or without regularisation)\footnote{We refer conventionally to the subsystems as `particles', but it should be borne in mind that the CV systems could equally be independent photon polarisation modes, photon modes or even joint photon and phonon degrees of freedom.}. One possibility is provided by the multipartite Greenberger-Horne-Zeilinger NOPA-like squeezed states \cite{vanLoockandBraunstein2001, Chen2002}. These have the virtue of experimental accessibility \cite{Ouetal1988, Ouetal1992}, and do show singular behaviour in the large squeezing limit which moreover leads to violations of the multipartite CHSH inequalities \cite{Kuzmichetal2000}. For a derivation of the CHSH inequalities for $N$-particle systems see for example \cite{Mermin1990}. However, many other choices of relative variables exist -- see for example \cite{Trifonov1998} and references therein.

In this paper we take up a logical generalisation of the original EPR suggestion, in selecting simultaneously diagonalisable joint degrees of freedom from amongst the canonical Jacobi relative coordinates of the particles. \S 2 is divided into two subsections. In the first we introduce the bipartite Fan and Klauder EPR-like state and propose a possible regularisation. This is shown to be identical to the bipartite NOPA state (which approximates the ideal EPR limit) with squeezing parameter $r\rightarrow \infty$. In fact the correspondence is that our regularisation $s \rightarrow 1^+$ coincides with $\tanh r \rightarrow 1^-$ as $r\rightarrow \infty$, with $\tanh r=1/s^2$. In the bipartite case the Fan and Klauder states are more general than the original EPR states in that they realise explicit nonzero eigenvalues of total or relative positions and momenta; but a study of the Wigner functions reveals that such nonzero eigenvalues can be absorbed into shifts of the complex displacement parameters, and so in the bipartite case the NOPA states do not lose any generality in not allowing for such nonzero eigenvectors. In the next subsection we follow the Fan-Klauder approach, in second quantised formalism, to derive the explicit theoretical tripartite CV states of EPR type conforming to this structure, and we develop the methods to provide a plausible regularisation. 

In \S 3, we derive explicit Wigner functions for our regularised tripartite states of EPR type by interpreting the Wigner functions themselves as expectation values of displaced parity measurement operators. As could be expected from their different second-quantised forms, the tripartite EPR and NOPA Wigner functions differ significantly (the appendix provides a comparison of the second-quantised forms and their respective Wigner functions, including an explicit evaluation of the former in the NOPA-like case, and a detailed derivation of the latter for our EPR-type states). Specifically, whereas the multipartite NOPA Wigner functions are singular in the large squeezing limit, our regularised tripartite states of EPR type admit two different singular regimes: not only in the EPR-type regime ($s \rightarrow 1$ in our notation), where of course the Wigner function still differs from that of NOPA, but also, rather unexpectedly, for an additional, auxiliary regime of the regulator ($s \rightarrow \sqrt{2}$). In \S 4 we exploit the fact that Wigner functions are immediately applicable as summands in the appropriate tripartite CHSH inequalities. We explore the two singular regimes and their Bell operator expectation values which control the classical-quantum boundary via the CHSH bound ($B_3 \le 2$) and identify some instances of violations for each of the cases $s\rightarrow 1^+$ and $s\rightarrow \sqrt{2^+}$. \S 5 includes further discussion of our findings, as well as comparison with the recent work \cite{FanandZhang1998, FanandLiu2007}, which provides a general construction of ideal EPR states, and some concluding remarks.

\section{Regularised CV EPR states}

\subsection{Bipartite states}\label{Bipsub}

The case considered by EPR in \cite{EPR1935} discusses the simultaneous diagonalisation of the two commuting variables of difference in position $(X_1-X_2)$ and total momentum $(P_1+P_2)$, where $X_j,\,P_j,\,\,j=1,2$ are a standard pair of canonically conjugate variables with $[X_j,P_k]=i\delta_{jk}$. Fan and Klauder \cite{FanandKlauder1994} give an explicit form for the common eigenvectors of the relative position and total momentum for two EPR particles in terms of creation operators as follows:
\begin{eqnarray}
|\eta\rangle = e^{-\frac 12 |\eta|{}^2 + \eta a{}^\dagger - \eta ^* b{}^\dagger +a{}^\dagger b{}^\dagger}|00\rangle,
\end{eqnarray}
where $\eta = \eta_1 + i\eta_2$ is an arbitrary complex number, $[a,a^\dagger]=1$, $[b,b^\dagger]=1$ and $|00\rangle \equiv |0,0\rangle$, the two-mode vacuum state. Thus 
\begin{eqnarray}
\left(X_1 - X_2\right)|\eta\rangle = \sqrt{2}\eta_1|\eta\rangle, \hspace{0.5cm} \left(P_1+P_2\right)|\eta\rangle=\sqrt{2}\eta_2|\eta\rangle,
\end{eqnarray}
with the coordinate and momentum operators definable as:
\begin{eqnarray}
X_1=\frac{1}{\sqrt{2}}\left(a+a^\dagger\right), \hspace{0.4cm} X_2=\frac{1}{\sqrt{2}}\left(b+b^\dagger\right), \hspace{0.4cm}  P_1=\frac{1}{i\sqrt{2}}\left(a-a^\dagger\right), \hspace{0.4cm} P_2=\frac{1}{i\sqrt{2}}\left(b-b^\dagger\right).
\end{eqnarray}

As a genuine representation of ideal generalised EPR states, with appropriate orthonormality and completeness, $|\eta\rangle$ is singular. In this paper we consider the following regularised version:
\begin{eqnarray}
|\eta\rangle_s &:=& N_2\,e^{-\frac{1}{2s^2} |\eta|^2 + \frac 1s \eta a^{\dagger} - \frac{1}{s} \eta^* b^{\dagger} + \frac{1}{s^2} a^{\dagger}b^{\dagger}} |00\rangle\label{Bieta},
\end{eqnarray}
with normalisation $|N_2|^2=|(s^4-1)|^{1/2}/s^2$.

The bipartite CV state (\ref{Bieta}) is to be compared with a regularised EPR-like state which has already appeared in the literature -- the so-called NOPA state from quantum optics \cite{Reid1989}-\cite{BanaszekandWodkiewicz1999}:
\begin{eqnarray}
|NOPA\rangle &=& e^{r\left( a^\dagger b^\dagger - a b\right)} |00\rangle.\label{NOPA}
\end{eqnarray}
NOPA states are produced by Nondegenerate Optical Parametric Amplification, and are the optical analog to the EPR state in the limit of strong squeezing. The NOPA state has already been shown to be a genuinely entangled state that produces violations of the CHSH inequality \cite{BanaszekandWodkiewicz1998, Ouetal1992, Kuzmichetal2000}. 

Following \cite{Yurkeetal1986} on reordering $SU(1,1)$ operators, we can reorder the expression for NOPA (\ref{NOPA}) into the following form:
\begin{eqnarray}
|NOPA\rangle &=& e^{r(a^{\dagger}b^{\dagger} - ab)}|00\rangle\nonumber\\
&=& e^{ra^{\dagger}b^{\dagger}}e^{-2\ln\,\cosh(r)\frac 12(a^{\dagger}a + b^{\dagger}b + 1)}e^{-rab}|00\rangle\nonumber\\
&=& \sqrt{1-\tanh{}^2 r}\;e^{\tanh ra^{\dagger}b^{\dagger}}|00\rangle.\label{RearrangedNOPA}
\end{eqnarray}
Note from (\ref{RearrangedNOPA}) that in the number basis, as $\tanh r\rightarrow1$ the $|NOPA\rangle$ state approximates the ideal EPR limit:
\begin{eqnarray}
\lim_{r\rightarrow\infty}|NOPA\rangle \approx  |EPR\rangle \approx |0,0\rangle + |1,1\rangle +|2,2\rangle +\ldots.
\end{eqnarray}

In \ref{etaappendix} it is argued that taking $\eta = 0$ corresponds to a shift in the parameters of the displacement operators (with some constraints on the choice of new parameters), such that we may rearrange the $|\eta\rangle_s$ regularisation to show that it approaches the NOPA regularisation, with $\tanh r = 1/s^2$. For $\eta = 0$ we therefore have:
\begin{eqnarray}
|\eta=0\rangle_s = N_{2}\,e^{\frac{1}{s^2}a^\dagger b^\dagger}|00\rangle\label{Bizeroeta}.
\end{eqnarray}

\subsection{Tripartite states}

A suitable analogue of equations (\ref{Bieta}) or (\ref{RearrangedNOPA}) which has the features required of an entangled state, which we analyse in detail below, is defined by:
\begin{eqnarray}
|\eta, \eta', \eta''\rangle_s= N_3 \,e^{-\frac{1}{4s^2} |\eta|^2-\frac{1}{4s^2} |\eta'|^2-\frac{1}{4s^2} |\eta''|^2 + \frac 1s (\eta a^{\dagger} + \eta' b^{\dagger} + \eta'' c^{\dagger} ) + \frac{1}{s^2} (a^{\dagger}b^{\dagger} + a^{\dagger}c^{\dagger} + b^{\dagger}c^{\dagger})} |000\rangle\label{Trieta},
\end{eqnarray}
with normalisation $|N_3|^2=|(s^4-1)^2\left(s^4-4\right)|^{1/2}/s^6$. For the case $\eta=\eta'=\eta'' = 0$, the tripartite EPR-like state becomes:
\begin{eqnarray}
|\eta=\eta'=\eta'' = 0\rangle_s= N_{3}e^{\frac{1}{s^2} (a^{\dagger}b^{\dagger} + a^{\dagger}c^{\dagger} + b^{\dagger}c^{\dagger})} |000\rangle.\label{trizeroeta}
\end{eqnarray}
Note here that, while the set of states (\ref{trizeroeta}) belong to the well known pure, fully symmetric three-mode Gaussian states, the more general case of (\ref{Trieta}) where the parameters $\eta$, $\eta'$ and $\eta''$ are retained is not symmetric, since the parameters can all differ. For discussion of Gaussian states in relation to entanglement in CV systems, see \cite{AdessoandIlluminatiREVIEW2007} and references therein.
 
Whereas the bipartite state $|\eta\rangle_{s}$ was a simultaneous eigenstate of $(X_1\!-\!X_2)$ and $(P_1\!+\!P_2)$, in the tripartite case the choice of relative variables is no longer immediately apparent. In a similar manner to the derivation of (\ref{Bieta}), it is readily established using manipulations of the type:
\begin{eqnarray}
ae^A = e^A\left\{a-[A,a]+\frac 12 [A,[A,a]]+\ldots\right\},
\end{eqnarray}
that generically $|\eta,\eta',\eta''\rangle_s$ is an eigenstate of the following combinations:
\begin{eqnarray}
\left( a-\frac{1}{s^2}\left(b^\dagger+c^\dagger\right)\right)|\eta,\eta',\eta''\rangle_s &=& \frac 1s \eta|\eta,\eta',\eta''\rangle_s,\nonumber\\
\left( b-\frac{1}{s^2}\left(c^\dagger+a^\dagger\right)\right)|\eta,\eta',\eta''\rangle_s &=& \frac 1s \eta'|\eta,\eta',\eta''\rangle_s,\nonumber\\
\left( c-\frac{1}{s^2}\left(a^\dagger+b^\dagger\right)\right)|\eta,\eta',\eta''\rangle_s &=& \frac 1s \eta''|\eta,\eta',\eta''\rangle_s.
\end{eqnarray}

From this it is clear that different values of $s$ will dictate limiting cases wherein $|\eta,\eta',\eta''\rangle_s$ becomes a singular eigenvalue of various choices of relative variables. (Note that in the bipartite case we could have introduced $|\eta\rangle_s$ as $|\eta,\eta'\rangle_s$ analogously, recovering (\ref{Bieta}) in the case $\eta'=-\eta^*$.) Keeping $s$ general, the eigenvalue equations become:
\begin{eqnarray}
\frac{1}{\sqrt{2}}\left(s\!+\!\frac 1s\right)(X_1\!-\!X_2)+\frac{i}{\sqrt{2}}\left(s\!-\!\frac 1s\right)(P_1\!-\!P_2)|\eta,\eta',\eta''\rangle_{s} &=& (\eta\!-\!\eta')|\eta,\eta',\eta''\rangle_{s},\nonumber\\
\frac{1}{\sqrt{2}}\left(s\!+\!\frac 1s\right)(X_2\!-\!X_3)+\frac{i}{\sqrt{2}}\left(s\!-\!\frac 1s\right)(P_2\!-\!P_3)|\eta,\eta',\eta''\rangle_{s} &=& (\eta'\!-\!\eta'')|\eta,\eta',\eta''\rangle_{s},\nonumber\\
\frac{1}{\sqrt{2}}\left(s\!-\!\frac 2s\right)(X_1\!+\!X_2\!+\!X_3)+\frac{i}{\sqrt{2}}\left(s\!+\!\frac 2s\right)(P_1\!+\!P_2\!+\!P_3)|\eta,\eta',\eta''\rangle_{s} &=& (\eta\!+\!\eta'\!+\!\eta'')|\eta,\eta',\eta''\rangle_{s}\label{NewEigen}.
\end{eqnarray}

From (\ref{NewEigen}) it is clear that the singular cases will occur for $s=1$ and $s=\sqrt{2}$. For the case $s=1$ we evidently have a singular eigenstate of the relative coordinates, while remaining a \emph{squeezed} state \cite{WallsandMilburn1994} of the total momentum. Conversely, for $s=\sqrt{2}$ we have a singular eigenstate of the total momentum, but a squeezed state of the relative coordinates. 

If we construct mode operators corresponding to the Jacobi relative variables and the canonical centre-of-mass variables, say
\begin{eqnarray}
\mathfrak{a}_{rel} &=& \frac 12\left(X_1-X_3\right)+\frac i2\left(P_1-P_3\right),\nonumber\\
\mathfrak{b}_{rel} &=& \frac{1}{2\sqrt{3}}\left(X_1+X_3-2X_2\right)+\frac{i}{2\sqrt{3}}\left(P_1+P_3-2P_2\right),\nonumber\\
\mathfrak{a}_{cm} &=& \frac{1}{\sqrt{6}}\left(X_1+X_2+X_3\right)+\frac{i}{\sqrt{6}}\left(P_1+P_2+P_3\right), 
\end{eqnarray}
then we find, from (\ref{NewEigen}) for general $s$:
\begin{eqnarray}
\left(s\mathfrak{a}_{rel} +\frac{1}{s}\mathfrak{a}^\dagger_{rel}\right)|\eta,\eta',\eta''\rangle_s &=& \frac{1}{\sqrt{2}} \left(\eta-\eta''\right)|\eta,\eta',\eta''\rangle_s,\nonumber\\
\left(s\mathfrak{b}_{rel} + \frac{1}{s} \mathfrak{b}^\dagger_{rel}\right)|\eta,\eta',\eta''\rangle_s &=& \frac{1}{\sqrt{6}} \left(\eta-2\eta'+\eta''\right)|\eta,\eta',\eta''\rangle_s,\nonumber\\
\left(s\mathfrak{a}_{cm} - \frac 2s \mathfrak{a}_{cm}^\dagger\right)|\eta,\eta',\eta''\rangle_s &=& \frac{1}{\sqrt{3}}\left(\eta+\eta'+\eta''\right)|\eta,\eta',\eta''\rangle_s,
\end{eqnarray}
from which it is again obvious that for $s=1$ or $s=\sqrt{2}$, canonical combinations arise in the first two, and last cases respectively. On the other hand, the non-canonical combinations appearing for $s=1$ in the third, and $s=\sqrt{2}$ in the first two cases, indicate that the squeezing parameters have the values $\frac 12 \ln 3$ in each instance.

Having established the structure of the tripartite EPR-like states, we can examine their behaviour when applied to Wigner functions, and the consequences of using these states in CHSH inequalities.

\section{Tripartite Wigner Function}\label{Wignersec}

The Wigner function \cite{Wigner1932, Hilleryetal1984}, was an attempt to provide the Schr\"odinger wavefunction with a probability in phase space. The time-independent function for one pair of $x$ and $p$ variables is:
\begin{eqnarray}
W(x,p) = \frac{1}{\pi \hbar} \int^{\infty}_{-\infty} dy \psi^* (x+y)\psi (x-y) e^{2ipy/\hbar}.\label{wignerfn}
\end{eqnarray}

Alternatively, it has been shown \cite{Moyal1949, Royer1977} that a useful expression of the Wigner function is in the form of quantum expectation values. For $N$ modes, the Wigner function for a state $|\psi\rangle$ may be expressed as the expectation value of the displaced parity operator, where the parity operator itself performs reflections about phase-space points ($\alpha_j)$, where $\alpha_j=\frac{1}{\sqrt{2}}(x_j+ip_j)$, with $j=1,2,\ldots,N$ denoting the mode, and:
\begin{eqnarray}
W(\alpha_1, \alpha_2, \ldots,\alpha_N) &=& \left(\textstyle{\frac{2}{\pi}}\right)^N\langle\Pi(\alpha_1, \alpha_2, \ldots,\alpha_N)\rangle.\label{Wigner}
\end{eqnarray}
The displaced parity operator is:
\begin{eqnarray}
\Pi(\alpha_1, \alpha_2, \ldots,\alpha_N) &=& {\otimes^{N}_{j=1}}D_j(\alpha_j)(-1)^{n_j}D_j^\dagger(\alpha_j),
\end{eqnarray}
where $n_j$ are the number operators, and for each mode the Glauber displacement operators are of the form:
\begin{eqnarray}
D(\alpha)=e^{\alpha a^\dagger - \alpha^*a}.
\end{eqnarray}

When we express the Wigner function in the form of (\ref{Wigner}), we can derive a set of these functions to construct the inequalities that will be discussed in section \ref{CHSHsect}. The tripartite Wigner function for $|\eta,\eta',\eta''\rangle_s$ becomes: 

\begin{eqnarray}
W(\alpha,\beta,\gamma) &=& \left(\frac{2}{\pi}\right)^3 \, N_3^2 \, e^{-\frac{1}{2s^2} |\eta|^2-\frac{1}{2s^2} |\eta'|^2-\frac{1}{2s^2} |\eta''|^2}\nonumber\\
&&\times \langle 000|\exp\left(\frac 1s \left(\eta^*a+\eta'^* b + \eta''^* c\right) + \frac{1}{s^2}\left(ab+ac+bc\right)\right)\nonumber\\
&&\times\; e^{\alpha a^\dagger - \alpha^*a}e^{\beta b^\dagger -\beta^*b}e^{\gamma c^\dagger - \gamma^* c}\left(-1\right)^{n_a+n_b+n_c}e^{\alpha^* a - \alpha a^\dagger}e^{\beta^*b -\beta b^\dagger}e^{\gamma^* c - \gamma c^\dagger}\nonumber\\
&&\times\; \exp\left(\frac 1s \left(\eta a^\dagger+\eta' b^\dagger + \eta'' c^\dagger\right) + \frac{1}{s^2}\left(a^\dagger b^\dagger+a^\dagger c^\dagger +b^\dagger c^\dagger \right)\right)|000\rangle\label{WignerStart}.
\end{eqnarray}

We evaluate such matrix elements by commuting mode operators with the parity operator and rearranging using BCH identities before casting the operators into anti-normal ordered form.
Then a complete set of coherent states is inserted and integrated over. As indicated in section \ref{Bipsub} and shown in \ref{etaappendix}, we can absorb the $\eta$, $\eta'$, $\eta''$ parameters by shifting the displacement parameters up to a factor: $W_{\eta,\eta',\eta''}(\alpha',\beta',\gamma')=E(\alpha',\beta',\gamma',\eta,\eta',\eta'')W_{0,0,0}(\alpha,\beta,\gamma)$. We are free to choose instances where $E(\alpha',\beta',\gamma',\eta,\eta',\eta'')=1$, and henceforth we assume $\eta=\eta'=\eta''=0$ unless otherwise stated, and write simply $W(\alpha,\beta,\gamma)$. With this shift in displacements, the Wigner function for our tripartite state becomes:
\begin{eqnarray}
W(\alpha,\beta,\gamma) &=& \frac{8}{\pi^3} \exp \left(\frac{1}{(s^4-4)(s^4-1)}\left[ C_1(|\alpha|^2 + |\beta|^2 + |\gamma|^2) \right.\right.\nonumber\\
&&+\;\; C_2(\alpha\beta + \alpha\gamma + \beta\gamma + \alpha^*\beta^* + \alpha^*\gamma^* + \beta^*\gamma^*) \nonumber\\
&&+\;\; C_3(\alpha\beta^* + \alpha\gamma^* + \beta\alpha^* + \beta\gamma^* + \gamma\alpha^* + \gamma\beta^*)\nonumber\\
&&+\;\; \left.C_4(\alpha^2 + \beta^2 + \gamma^2 + \alpha^{*2} + \beta^{*2} + \gamma^{*2}) \right]\Bigg)\label{Wigner2},
\end{eqnarray}
where
\begin{eqnarray}
C_1=-2(s^8-s^4-4)\,,\hspace{0.25cm} C_2=4s^2(s^4-2)\,,\hspace{0.25cm} C_3=-4s^4\,,\hspace{0.25cm} C_4=4s^2.\nonumber
\end{eqnarray}

The most important point to note here is the emergence of mixed conjugate/non-conjugate pairs, which do not appear in the Wigner function for the second-quantised NOPA-like optical analogue (see \ref{A1}). To make the behaviour of the Wigner function in the asymptotic region clearer, the parameters $\alpha$, $\beta$ and $\gamma$ are written in polar form, $\alpha = |\alpha|e^{i\phi_\alpha}$ etc. The Wigner function thus becomes:

\begin{eqnarray}
W(\alpha,\beta,\gamma) &=&\frac{8}{\pi^3}\exp\left(\frac{1}{(s^4-4)(s^4-1)}\left[ C_1\left(|\alpha|^2+|\beta|^2+|\gamma|^2\right)\right.\right.\nonumber\\
&& + \;2C_2\left(|\alpha||\beta|\cos(\phi_\alpha+\phi_\beta) +|\beta||\gamma|\cos(\phi_\beta+\phi_\gamma) +  |\gamma||\alpha|\cos(\phi_\gamma+\phi_\alpha)\right)\nonumber\\
&&+ \;2C_3\left(|\alpha||\beta|\cos(\phi_\beta-\phi_\alpha) +
|\beta||\gamma|\cos(\phi_\gamma-\phi_\beta) + |\gamma||\alpha|\cos(\phi_\gamma-\phi_\alpha)\right)\nonumber\\
&&\left. + \;2C_4\left(|\alpha|^2\cos(2\phi_\alpha) + |\beta|^2\cos(2\phi_\beta)+ |\gamma|^2\cos(2\phi_\gamma)\right)\right]\Bigg)\label{PolarWigner}.
\end{eqnarray}

\section{CHSH inequalities and violations}\label{CHSHsect}

The CHSH inequalities \cite{Clauseretal1969} are CV generalisations of the original Bell inequalities which were set up to test the scheme proposed by Einstein, Podolsky and Rosen (EPR) in 1935 \cite{EPR1935}. Considering the measurement of an entangled pair of particles performed after they have been separated such that no classical communication channels are open when the wavefunction collapses, EPR posited that either quantum mechanics must be incomplete, with room for a hidden variable theory, or spatiotemporal locality is violated. Bell showed that hidden variables were not permitted if we preserve both the assumptions of accepted theory -- specifically locality -- and the probabilities predicted by quantum mechanics. 

Generalised $N$-mode Bell inequalities -- CHSH inequalities, in terms of the Bell operator expectation values $B_N$ -- exist (see for example \cite{Mermin1990}). In their bi- and tripartite form we can apply these to our regularised EPR-like states. Following \cite{BanaszekandWodkiewicz1998, vanLoockandBraunstein2001}, the CHSH inequalities for the bi- and tripartite forms are the possible combinations: 
\begin{eqnarray}
B_2 &=& \Pi(0,0) + \Pi(0,\beta) + \Pi(\alpha,0) - \Pi(\alpha,\beta),\\
|B_2| &\leq& 2,\nonumber\\
B_3 &=& \Pi(0,0,\gamma) + \Pi(0,\beta,0) + \Pi(\alpha,0,0) - \Pi(\alpha,\beta,\gamma),\label{B3}\\
|B_3| &\leq& 2.\nonumber
\end{eqnarray}

\begin{figure}[!ht]
\centering
\includegraphics[width=10cm]{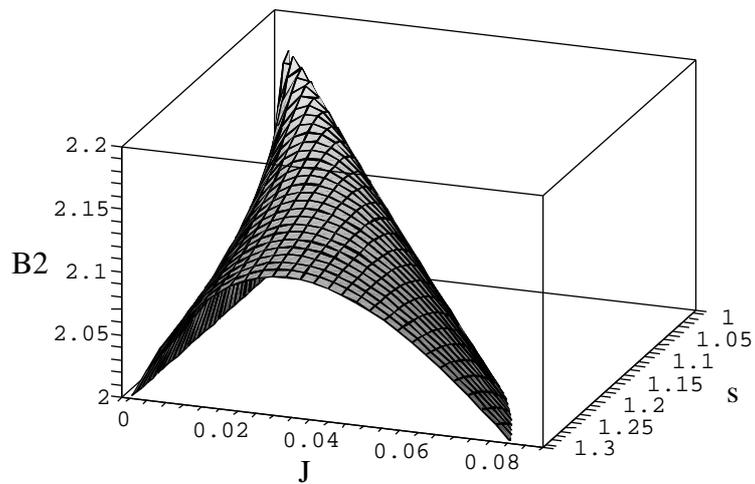}
 \caption{Plot of bipartite $s$-modified CHSH, with an all-imaginary choice for $\alpha$ and $\beta$. Reaches a maximum value of $\approx 2.19$ as $s\rightarrow 1$ and $J\rightarrow 0$. Note that this is equivalent to the NOPA case.}
\label{Bipartite}
\end{figure}

In \cite{vanLoockandBraunstein2001}, the CHSH inequality constructed with the tripartite NOPA-like state (see equation (\ref{NOPA3})) is maximised by taking an all-imaginary substitution $\alpha =\beta=\gamma=i\sqrt{J}$, where $J$ is some distance measure. In the bipartite case (Figure \ref{Bipartite}), if we look in the region $s\rightarrow 1^+$ for (\ref{Bizeroeta}), an all-imaginary substitution obviously gives exactly the same maximum violation as NOPA with $r\rightarrow \infty$, both having a maximum value of $B_2^{max} \approx 2.19$ \cite{BanaszekandWodkiewicz1998, vanLoockandBraunstein2001}. The figure shows clearly that the value of $B_2$ increases as $s\rightarrow 1^+$ and $J\rightarrow 0$.

In the tripartite case however, we must examine the wealth of other possible choices which extremise the inequality. From the form of the Wigner function in (\ref{PolarWigner}), however, there are some clear choices that will minimise the last term in $B_3$. Choosing all the phases $\phi_\alpha=\phi_\beta=\phi_\gamma=\frac{\pi}{2}$, and all the magnitudes $|\alpha|=|\beta|=|\gamma|=\sqrt{J}$, such that all the parameters are imaginary $(i\sqrt{J})$, equation (\ref{PolarWigner}) becomes:
\begin{eqnarray}
W(i\sqrt{J},0,0)&=&W(0,i\sqrt{J},0)=W(0,0,i\sqrt{J})= \frac{8}{\pi^3}\exp\left(-\frac{J(s^4-s^2+2)}{(s^2+1)(s^2-2)}\right),\nonumber\\
W(i\sqrt{J},i\sqrt{J},i\sqrt{J}) &=& \frac{8}{\pi^3}\exp\left(-\frac{3J(s^2+2)}{s^2-2}\right).
\end{eqnarray}
Consequently $B_3$ is (from (\ref{B3})):
\begin{eqnarray}
B_3=3\exp\left\{-\frac{J(s^4-s^2+2)}{(s^2+1)(s^2-2)}\right\}-\exp\left\{-\frac{3J(s^2+2)}{(s^2-2)}\right\}\label{ImB3}.
\end{eqnarray}

\begin{figure}[!ht]
\centering
\includegraphics[width=10cm]{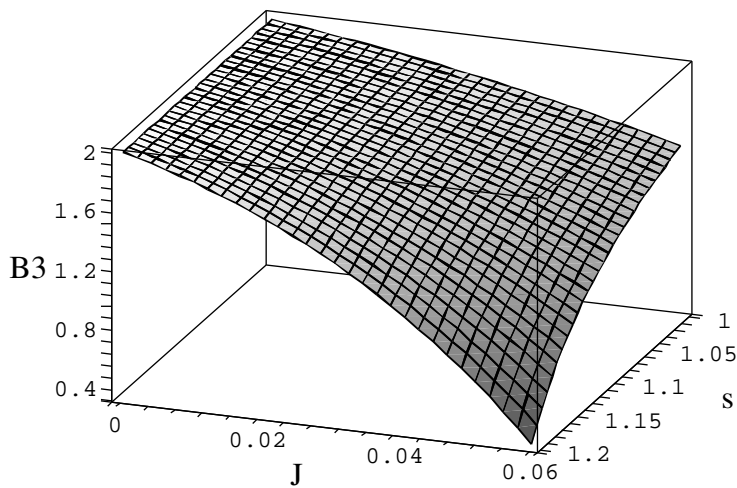}
 \caption{Tripartite $s$-modified CHSH. With an all-imaginary choice for $\alpha$, $\beta$ and $\gamma$, $B_3$ never reaches a value greater than 2 as $s\rightarrow 1$.}
\label{Tripartite2}
\end{figure}

\begin{figure}[!ht]
\centering
\includegraphics[width=10cm]{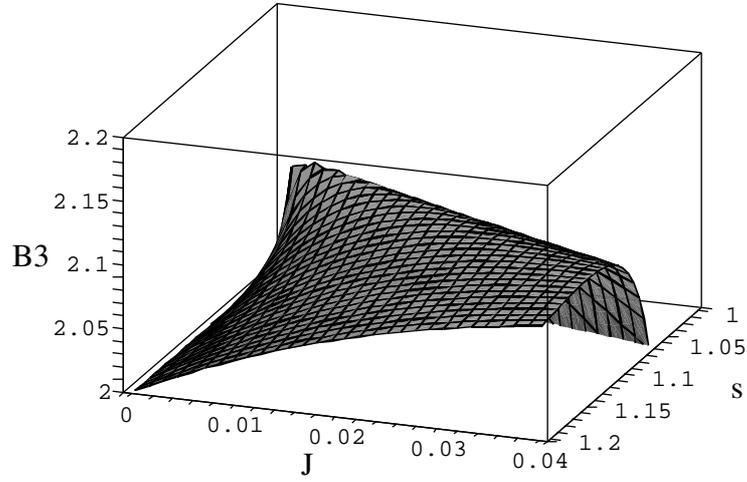}
 \caption{Tripartite $s$-modified CHSH. With $\alpha=-\beta=-\sqrt{J}$, $\gamma=0$, $B_3$ reaches a maximum value of $\approx 2.09$ as $s\rightarrow 1^+$ and $J\rightarrow 0$.}
\label{Tripartite2.09}
\end{figure}

In the region $s\rightarrow 1^+$, $B_3$ never reaches a value greater than  $2$ (Figure \ref{Tripartite2}). A violation corresponding to the EPR limit $s\rightarrow1^+$ can be found by making the choice $\alpha = -\beta = -\sqrt{J}$; $\gamma=0$, for which (\ref{PolarWigner}) gives:
\begin{eqnarray}
W(-\sqrt{J},0,0)&=&W(0,\sqrt{J},0) = \exp\left\{-\frac {J \left( {s}^{4}+{s}^{2}+2 \right) }{ \left( s^2-1 \right)  \left( s^2+2 \right) }\right\},\nonumber\\
W(-\sqrt{J},\sqrt{J},0) &=& \frac{8}{\pi^3}\exp\left(-\frac{2J(s^2+1)}{(s-1)(s+1)}\right),\nonumber\\
W(0,0,0)&=&1,
\end{eqnarray}
and $B_3$ becomes (Figure \ref{Tripartite2.09}):
\begin{eqnarray}
B_3&=& 1+2\,\exp\left\{-\frac {J \left( {s}^{4}+{s}^{2}+2 \right) }{ \left( s^2-1 \right)  \left( s^2+2 \right) }\right\}
 -\exp\left\{-\frac {2 J \left( {s}^{2}+1 \right)}{ \left( s^2-1\right) }\right\}.
\end{eqnarray}
As $s\rightarrow 1^+$, $J\rightarrow 0$, the maximum value is $B_3^{max} \approx 2.09$, which can be checked both analytically and numerically.

\begin{figure}[!ht]
\centering
\includegraphics[width=10cm]{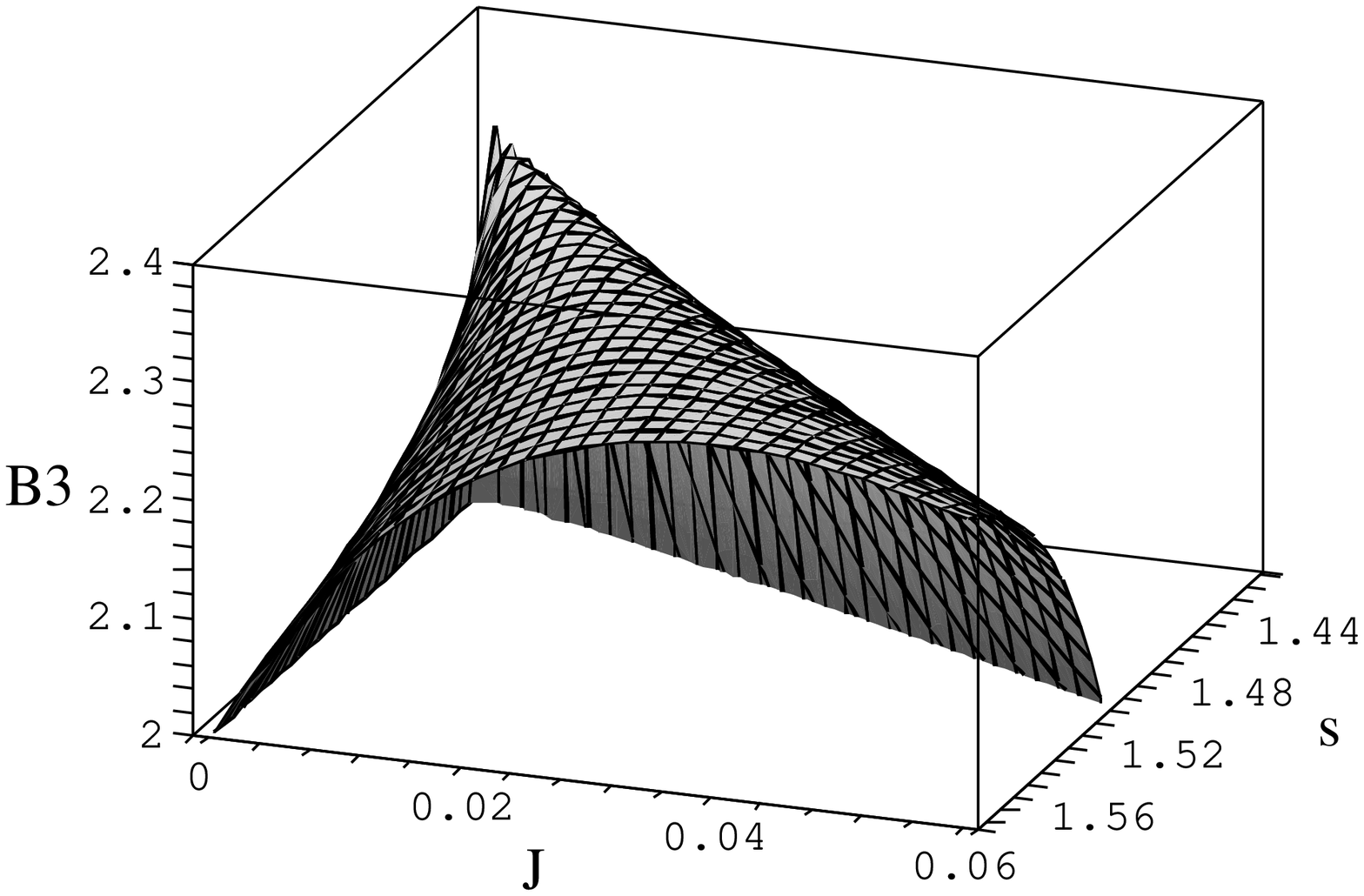}
 \caption{Tripartite $s$-modified CHSH. With an all imaginary choice for $\alpha$, $\beta$ and $\gamma$, $B_3$ reaches a maximum value of $\approx 2.32$ as $s\rightarrow \sqrt{2^+}$ and $J\rightarrow 0$.}
\label{Tripartite2.32}
\end{figure}

However, what is more interesting still is exploring an auxiliary regime of the regulator, $s\rightarrow \sqrt{2^+}$, in equation (\ref{ImB3}). This is shown in Figure \ref{Tripartite2.32}. Analytically, we can approximate the maximum to the lowest order in $s-\sqrt{2}=\epsilon$ by writing $B_3=3x-x^\lambda$, where $x=\exp(-4J/3\epsilon^2)$, and $\lambda=9$, with maximum
\begin{eqnarray}
B_3^{max} &\cong& (\lambda-1)\left(\frac{3}{\lambda}\right)^{\frac{\lambda}{\lambda-1}} \cong 2.32.\label{B3max}
\end{eqnarray}
at $x=\left(\frac{3}{\lambda}\right)^{\frac{1}{\lambda-1}}$. This can be confirmed numerically for $s\rightarrow \sqrt{2^+}$, $J\rightarrow 0$. The values of $B_3^{max}$ correspond exactly to those calculated for the experimentally verified NOPA-like states, whose maximisation as $r \rightarrow \infty$ is also governed by (\ref{B3max}).

\section{Discussion}

In this paper we have analysed tripartite CV entangled states which are natural generalisations of the classic bipartite EPR-type states (for two systems with canonical variables $X_1$, $P_1$, $X_2$, $P_2$). Given the necessity of working with normalisable states which still approximate the ideal EPR-type limit for practical implementation of CHSH inequalities, we examined a family of such regulated states parameterized by a regulating parameter $s$. This family of states was compared with those relating to multipartite NOPA-like states. The NOPA states have been shown to manifest CHSH violations, and have the advantage of being directly accessible by experiment via standard quantum optics protocols such as multiparametric heterodyne detection techniques and beam splitter operations. However, as an extension of a direct transcription of the EPR paradox, this new family of regularised states provides an alternative, systematic description of the approach to the ideal EPR states for relative variables.

By finding expressions for the eigenstates of the regularised tripartite CV EPR-like states it became apparent that there are two regimes of the regularisation parameter in which these states become singular: in one case ($s \rightarrow 1$) we have a singular eigenstate of the relative coordinates while remaining squeezed in the total momentum; in the other, $s\rightarrow \sqrt{2}$ limit we have a singular eigenstate of the total momentum, but squeezed in the relative coordinates. In these two regimes we have explored CHSH inequalities via Wigner functions regarded as expectation values of displaced parity operators. Violations of the tripartite CHSH bound ($B_3 \le 2$) are established analytically and numerically, with $B_3 \cong 2.09$ in the canonical regime ($s \rightarrow 1^+$), as well as $B_3 \cong 2.32$ in the auxiliary regime ($s \rightarrow \sqrt{2^+}$).

Related tripartite entangled states have recently been constructed by Fan \cite{Fan2006}. Although these states are also accessible by standard quantum optics techniques, they are not true generalisations of `EPR' states. In this case, while they diagonalise one centre-of-mass variable (for example, $X_1\!+\!X_2\!+\!X_3$), they are \emph{coherent} states \cite{KlauderandSkagerstam1985} of the remaining relative Jacobi observables (that is, they diagonalise their annihilation mode operators ${\mathfrak a}$, ${\mathfrak b}$ in contrast to the $s\rightarrow \sqrt{2}$ limit of our EPR-type tripartite states, which as stated above turn out to be \emph{squeezed} states of these relative degrees of freedom (eigenstates of a linear combination $\frac{1}{\sqrt{3}}(2{\mathfrak a} + {\mathfrak a}^\dagger)$, $\frac{1}{\sqrt{3}}(2{\mathfrak b} + {\mathfrak b}^\dagger)$ in the relative mode operators, with the value $\frac 12 \ln 3$ for the squeezing parameter). In the case of the tripartite entangled states of \cite{Fan2006}, no regularisation has been given. A construction of true multipartite ideal EPR states has, however, been provided in \cite{FanandZhang1998, FanandLiu2007}, with a second-quantised form for the tripartite state (\ref{Fan3}), which may be compared with the form for the NOPA-like state (\ref{QuantizedNOPA3}). Although the Wigner functions for the tripartite NOPA-like states show peaks at zeroes of $X_i-X_j$ and $P_i+P_j$ for all distinct pairs $i,j$ \cite{BanaszekandWodkiewicz1999}, which does not appear to be consistent with simultaneous diagonalisation of commuting observables, it can be inferred from the agreement of (\ref{Fan3}) and (\ref{QuantizedNOPA3}) that indeed in the infinite squeezing limit, $\tanh(r)=1$, and with relative parameters equal to zero, the NOPA-like state does again tend to the ideal EPR state.

Since the NOPA-like states are constructed with a view to experimental realisability, and, in the bipartite case, to manufacture the specific properties of the Wigner function, this new suggestion for a regularisation stemming from a direct transcription of the EPR paradox in terms of the simultaneous diagonalisation of commuting observables could be seen as a more general or fundamental description. It also considers in more detail the specific instance of tripartite EPR-type states, compared to the comprehensive \cite{FanandLiu2007} which finds $n$-partite representations of entangled states through their Gaussian-form completeness relation without exploring regularisations and Wigner function properties. As the proposed EPR-type regularised state produces a different Wigner function from the NOPA-type, with two singular limits, this paper's proposed regularisation may potentially suggest that alternative experimental ways to achieve the violations of the CHSH inequalities are possible. For a review of Gaussian states, and discussions of the realisability of entangled states, we refer to \cite{AdessoandIlluminatiREVIEW2007, AdessoandIlluminati2005, FerraroandParis2005}, and references therein. It will be worth investigating the full extent of the constraints placed on the choices of displacement parameters entailed by the shift in $\eta$ (see \ref{etaappendix}). The current discussion might also easily be extended to include a presentation of the alternative bipartite starting point of conjugate variable choice $X_1+X_2$ and $P_1-P_2$. \cite{Trifonov1998} discusses the canonical combinations for any number of modes, but in our case it is reasonable to assume that an $N$-partite generalisation would be of the form $\left[\exp\left(\frac{1}{s^2}\left(\sum_{i<j}a_i^\dagger a_j^\dagger\right)\right)\right]$. We would also expect that these states would admit standard completeness relations in the singular cases.

In conclusion, we have presented a rigorous extension of Fan and Klauder's general EPR-like states to the regularised tripartite CV case for relative variables, and highlighted the connection to current quantum optics implementations. The CHSH inequalities constructed with component Wigner functions for this case show significant violation of the classical bound, and the different choices of regularisation parameter making the state singular illustrate an interesting new feature of the structure of generalised CV EPR-like states.

\section*{Acknowledgements}
This research was partially supported by the Commonwealth of Australia through the International Endeavour Awards. We thank Robert Delbourgo for a careful reading of the manuscript, and the referees for their suggestions helping to improve the presentation of this work.

\pagebreak

\pagebreak
\begin{appendix}
\section*{Deriving the tripartite NOPA and EPR-like Wigner functions}
\setcounter{section}{1}

\noindent In this appendix the tripartite NOPA and $|\eta, \eta', \eta''\rangle_s$ Wigner functions are derived for comparison.

\subsection{Second-quantised form and Wigner function of tripartite NOPA-like state}\label{A1}
Applying two phase-free beamsplitters at specified angles acting on one momentum squeezed state and two position squeezed states of mode $1$, $2$ and $3$ respectively, \cite{vanLoockandBraunstein2000} states that the tripartite NOPA-like states can be derived from the following expression:

\begin{eqnarray}
|NOPA^{(3)}\rangle &=& B_{23}\left(\frac{\pi}{4}\right) B_{12}\left(\rm{arccos}\frac{1}{\sqrt{3}}\right)\nonumber\\  &&\times\exp\left(\frac{r}{2}\left(a^2\!-\!a^{\dagger 2}\right)\right) \exp\left(\frac{-r}{2}\left(b^2\!-\!b^{\dagger 2}\right)\right) \exp\left(\frac{-r}{2}\left(c^2\!-\!c^{\dagger 2}\right)\right)|000\rangle.
\end{eqnarray}

We can therefore use the following formula quoted in \cite{Truax1985}\footnote{Note the misprint in the sign of the last exponential in \cite{Truax1985}; see \cite{FisherNietoandSandberg1984}} for the squeezing operator $S(z)$ (where $z=e^{i\theta}$) with the Baker-Campbell-Hausdorff (BCH) relation:

\begin{eqnarray}
S(z) &=& \exp \left[\frac 12 (za^{\dagger 2} - \overline{z} a^2)\right]\nonumber\\
&=& \exp\left[\frac 12 (e^{i\theta}\tanh r)a^{\dagger 2}\right] \exp\left[-2(\ln\,\cosh r)(\frac 12 a^{\dagger}a \!+ \!\frac 14)\right]\exp \left[-\frac 12 (e^{-i\theta}\tanh r)a^2\right].\label{BCH}
\end{eqnarray}

The following beamsplitter operation\footnote{An additional overall relative sign ($180^\circ$ phase shift) between the two modes has been omitted; see for example \cite{Hamilton2000}} can then be applied, where $\theta$ here refers to the angles $\pi/4$ and $\arccos(1/\sqrt{3})$ for the $B_{23}$ and $B_{12}$ splitters respectively:
\begin{eqnarray}
B_{ab}(\theta) : \left\{\begin{array}{c} a \rightarrow a \cos\theta + b\sin\theta \\  b \rightarrow - a\sin\theta + b\cos\theta \end{array} \right.
\end{eqnarray}
and normalising, the tripartite NOPA is expressible in second-quantised form as: 
\begin{eqnarray}
|NOPA^{(3)}\rangle &=& \left(1-\tanh^2(r)\right)^{3/4}\nonumber\\
&& \times\exp\left(-\frac 16\tanh r\left(  a^{\dagger 2}\! + \! b^{\dagger 2} \!+ \!c^{\dagger 2}\right)  + \frac 23 \tanh r\left(b^{\dagger}c^{\dagger}\! + \! a^{\dagger}b^{\dagger}\! +\!  a^{\dagger}c^{\dagger}\right)\right)|000\rangle\label{QuantizedNOPA3}.
\end{eqnarray}

Using this state to derive the Wigner function of the form (\ref{Wigner}), the complex exponential that is produced may be rearranged using the formula (\ref{Berezin}). The tripartite NOPA Wigner function then becomes:
\begin{eqnarray}
W_{NOPA} &=&\left(\frac{2}{\pi}\right)^3  \exp\left\{\left(2-\frac{4}{1-\tanh{}^2 r}\right)\left(|\alpha|^2+|\beta|^2+|\gamma|^2\right)\right.\nonumber\\
&& -\frac {2\tanh r}{3\left(1-\tanh{}^2 r\right)} \left(\alpha^{*2}+\beta^{*2}+\gamma^{*2}+\alpha^2+\beta^2+\gamma^2\right) \nonumber\\
&&\left.+ \frac{8\tanh r}{3\left(1-\tanh{}^2 r\right)}\left(\alpha\beta + \beta\gamma + \gamma\alpha + \alpha^*\beta^* + \beta^*\gamma^* + \gamma^*\alpha^*\right)\right\}\nonumber\\
&=& \frac{8}{\pi^3}  \exp\left\{\left(-2\cosh(2r)\right)\left(|\alpha|^2+|\beta|^2+|\gamma|^2\right)\right.\nonumber\\
&& -\frac 13 \sinh(2r) \left(\alpha^{*2}+\beta^{*2}+\gamma^{*2}+\alpha^2+\beta^2+\gamma^2\right)\nonumber\\
&&\left. + \frac 43 \sinh(2r)\left(\alpha\beta + \beta\gamma + \gamma\alpha + \alpha^*\beta^* + \beta^*\gamma^* + \gamma^*\alpha^*\right) \right\}.\label{NOPA3}
\end{eqnarray}
This is the result quoted in \cite{vanLoockandBraunstein2001}, and further explication can be found in that paper. This function should be compared with the Wigner function of our regularised tripartite EPR-like state, $|\eta,\eta',\eta''\rangle_s$. Further details of that derivation are given below.

\subsection{Derivation of conditions for $\eta=0$}\label{etaappendix}

In the interest of brevity, the conditions for shifting the $\eta$ parameters are shown below for the biparite case. However, the analysis extends in an obvious way to the triparite case. After BCH and anti-normal ordering, the bipartite Wigner function (tripartite given in equation (\ref{WignerStart})) becomes:
\begin{eqnarray}
W(\alpha, \beta) &=& _s\langle \eta,\eta'|\underbrace{e^{2|\alpha|^2}e^{2|\beta|^2}e^{-2\alpha^*a}e^{-2\beta^*b}e^{2\alpha a^\dagger}e^{2\beta b^\dagger}}_{F(\alpha,\beta)}(-1)^{n_a+n_b}|\eta,\eta'\rangle_s\nonumber\\
&=& \langle 00|\exp\left(-\frac {1}{4s^2}|\eta|^2 -\frac {1}{4s^2}|\eta'|^2 +\frac 1s\eta^*a +\frac 1s\eta'^*b +\frac {1}{s^2}ab\right) F(\alpha,\beta)\nonumber\\
&&\times\exp\left(-\frac {1}{4s^2}|\eta|^2 -\frac {1}{4s^2}|\eta'|^2 -\frac 1s\eta a^\dagger -\frac 1s\eta' b^\dagger +\frac {1}{s^2}a^\dagger b^\dagger\right)|00\rangle.
\end{eqnarray}
We then make the generic substitutions
\begin{eqnarray}
\alpha &=& \alpha' + A(\eta,s),\nonumber\\
\beta &=& \beta' + B(\eta',s),
\end{eqnarray}
into $F(\alpha,\beta)$. To find the expressions for $A(\eta,s)$ and $B(\eta',s)$ that will allow us to set $\eta=\eta'(=\eta'')=0$, we solve the following:
\begin{eqnarray}
2\alpha'A^* + 2\alpha'^*A+2|A|^2 - 2A^*a+2Aa^\dagger &=& \frac {1}{2s^2} |\eta|^2 - \frac 1s \eta^*a + \frac 1s \eta a^\dagger,\nonumber\\
2\beta'B^* + 2\beta'^*B+2|B|^2 - 2B^*b+2Bb^\dagger &=& \frac {1}{2s^2} |\eta'|^2 - \frac 1s \eta'^*b + \frac 1s \eta' b^\dagger.
\end{eqnarray}
Thus we can see that, if we allow the constraints 
\begin{eqnarray}
\frac{\alpha'\eta^*}{s}+\frac{\alpha'^*\eta}{s}&=&0,\nonumber\\
\frac{\beta'\eta'^*}{s}+\frac{\beta'^*\eta'}{s}&=&0,
\end{eqnarray}
(i.e. $\alpha'$ real and $\eta$ imaginary or vice versa), then the expressions for $A(\eta,s)$ and $B(\eta',s)$ become
\begin{eqnarray}
A(\eta,s)=\frac{\eta}{2s}, \hspace{0.5cm} A^*(\eta,s)=\frac{\eta^*}{2s},\nonumber\\
B(\eta',s)=\frac{\eta'}{2s}, \hspace{0.5cm} B^*(\eta',s)=\frac{\eta'^*}{2s}.
\end{eqnarray}

Therefore, up to a factor, taking $\eta=\eta'(=\eta'')=0$ corresponds to a shift in the parameters of the displacement operators $\alpha'=\alpha-\frac{\eta}{2s}$, $\beta'=\beta-\frac{\eta'}{2s}$ (and $\gamma'=\gamma-\frac{\eta''}{2s}$ in the tripartite case). For the tripartite Wigner function as used in section (\ref{Wignersec}), this can be expressed as:
\begin{eqnarray} W_{\eta,\eta',\eta''}(\alpha',\beta',\gamma')&=&E(\alpha',\beta',\gamma',\eta,\eta',\eta'')W_{0,0,0}(\alpha,\beta,\gamma)\\
E(\alpha',\beta',\gamma',\eta,\eta',\eta'')&=&\exp\left(\frac 1s(\alpha'\eta^* +\alpha'^*\eta +\beta'\eta'^* +\beta'^*\eta' +\gamma'\eta''^* +\gamma'^*\eta'')\right)
\end{eqnarray}

\subsection{Details of derivation of tripartite $|\eta,\eta',\eta''\rangle_s$ Wigner function}\label{A2}

The second-quantised EPR-like state is expressed as (\ref{Trieta}). This is used to find the Wigner function in the form of (\ref{WignerStart}). By commuting mode operators with the parity operator and rearranging using BCH identities, the expression becomes, in anti-normal ordered form:
\begin{eqnarray}
W &=& \left(\frac{2}{\pi}\right)^3 \, N_3^2 \, e^{-\frac{1}{2s^2} |\eta|^2-\frac{1}{2s^2} |\eta'|^2-\frac{1}{2s^2} |\eta''|^2}\nonumber\\
&&\times \langle 000|\exp\left(\frac 1s \left(\eta^*a+\eta'^* b + \eta''^* c\right) + \frac{1}{s^2}\left(ab+ac+bc\right)\right)\nonumber\\
&&\times\; e^{2|\alpha|^2}e^{2|\beta|^2}e^{2|\gamma|^2}e^{-2\alpha^*a}e^{-2\beta^*b}e^{-2\gamma^*c}e^{2\alpha a^\dagger}e^{2\beta b^\dagger}e^{2\gamma c^\dagger}\nonumber\\
&&\times\; \exp\left(\frac 1s \left(\eta a^\dagger+\eta' b^\dagger + \eta'' c^\dagger\right) + \frac{1}{s^2}\left(a^\dagger b^\dagger+a^\dagger c^\dagger +b^\dagger c^\dagger \right)\right)|000\rangle.
\end{eqnarray}
In anti-normal ordered form we may insert a complete set of coherent states $\int |u,v,w\rangle\langle u,v,w|\frac{d^2ud^2vd^2w}{\pi}$ such that we may rearrange the exponential according to the formula \cite{Berezin1966, Fan1990}:
\begin{eqnarray}
\int \prod^{n}_{i}\left[ \frac {d^2 z_{i}}{\pi}\right] \exp\left(-\frac 12 (z,z^*)\left(\begin{array}{cc}A & B \\ C & D \end{array}\right)\left(\begin{array}{c} z\\z^* \end{array}\right) + (\mu, \nu^*)\left(\begin{array}{c} z\\z^* \end{array}\right)\right)\nonumber\\
= \left[det\left(\begin{array}{cc}C & D \\ A & B \end{array}\right)\right]^{-\frac 12} \exp\left[\frac 12(\mu,\nu^*)\left(\begin{array}{cc}A & B \\ C & D \end{array}\right)^{-1} \left(\begin{array}{c} \mu\\\nu^* \end{array}\right)\right]\nonumber\\
= \left[det\left(\begin{array}{cc}C & D \\ A & B \end{array}\right)\right]^{-\frac 12} \exp\left[\frac 12(\mu,\nu^*)\left(\begin{array}{cc}C & D \\ A & B \end{array}\right)^{-1} \left(\begin{array}{c} \nu^*\\\mu \end{array}\right)\right]\label{Berezin},
\end{eqnarray}
where matrices $A$ and $D$ must be symmetrical, and $C=B^T$. In this instance
\begin{eqnarray}
(z,z^*)&=& \left(u,v,w,u^*,v^*,w^*\right),\nonumber\\
(\mu,\nu^*)&=&\left(\frac 1s \eta^*-2\alpha^*,\frac 1s \eta'^*-2\beta^*,\frac 1s \eta''^*-2\gamma^*,-\frac 1s \eta+2\alpha,-\frac 1s \eta'+2\beta,-\frac 1s \eta''+2\gamma \right),
\end{eqnarray}
and we have
\begin{eqnarray}
\left(\begin{array}{cc}C & D \\ A & B \end{array}\right)= \left(\begin{array}{cccccc} 1&0&0&0&-\frac{1}{s^2}&-\frac{1}{s^2}\\0&1&0&-\frac{1}{s^2}&0&-\frac{1}{s^2}\\0&0&1&-\frac{1}{s^2}&-\frac{1}{s^2}&0\\0&-\frac{1}{s^2}&-\frac{1}{s^2}&1&0&0\\-\frac{1}{s^2}&0&-\frac{1}{s^2}&0&1&0\\-\frac{1}{s^2}&-\frac{1}{s^2}&0&0&0&1 \end{array}\right),
\end{eqnarray}
with inverse
\begin{eqnarray}
\left(\begin{array}{cc}C & D \\ A & B \end{array}\right)^{-1} = \frac{s^4}{\left(s^4 -4\right)\left(s^4-1\right)}\left( \begin {array}{cccccc} {s}^{4}-3&1&1&2\,{s}^{-2}&{\frac {{s}^{
4}-2}{{s}^{2}}}&{\frac {{s}^{4}-2}{{s}^{2}}}\\\noalign{\medskip}1&{s}^
{4}-3&1&{\frac {{s}^{4}-2}{{s}^{2}}}&2\,{s}^{-2}&{\frac {{s}^{4}-2}{{s
}^{2}}}\\\noalign{\medskip}1&1&{s}^{4}-3&{\frac {{s}^{4}-2}{{s}^{2}}}&
{\frac {{s}^{4}-2}{{s}^{2}}}&2\,{s}^{-2}\\\noalign{\medskip}2\,{s}^{-2
}&{\frac {{s}^{4}-2}{{s}^{2}}}&{\frac {{s}^{4}-2}{{s}^{2}}}&{s}^{4}-3&
1&1\\\noalign{\medskip}{\frac {{s}^{4}-2}{{s}^{2}}}&2\,{s}^{-2}&{
\frac {{s}^{4}-2}{{s}^{2}}}&1&{s}^{4}-3&1\\\noalign{\medskip}{\frac {{
s}^{4}-2}{{s}^{2}}}&{\frac {{s}^{4}-2}{{s}^{2}}}&2\,{s}^{-2}&1&1&{s}^{
4}-3\end {array} \right).
\end{eqnarray}
Note also that
\begin{eqnarray}
\left[\rm{det}\left(\begin{array}{cc}C & D \\ A & B \end{array}\right)\right]^{-\frac 12} = \left[(s^{12}-6s^7+9s^4-4)/s^{12}\right]^{-\frac 12} = \frac{1}{N_3^2},
\end{eqnarray}
such that the $N_3^2$ cancel in the Wigner function.

From the argument in \ref{etaappendix}, we assume that $\eta=\eta'=\eta''=0$ unless otherwise specified, and continue to use $W(\alpha,\beta,\gamma)$. This gives equation (\ref{Wigner2}), which may now easily be compared with the Wigner function derived for the NOPA-like case (equation (\ref{NOPA3})). Further discussion of similar manipulations of the Wigner function can be found in \cite{deGosson2004}.

\subsection{Tripartite entangled state from \cite{FanandLiu2007}}
Equation $(27)$ in \cite{FanandLiu2007} provides the ideal EPR state for the tripartite entangled state:
\begin{eqnarray}
|p,\xi_2,\xi_3\rangle &=& \frac{1}{\sqrt{3}\pi^{\frac 34}}\exp\left[A+\frac{i\sqrt{2}p}{3}\sum^{3}_{i=1}a_i^\dagger+\frac{\sqrt{2}\xi_2}{3}(a_1^\dagger-2a_2^\dagger +a_3^\dagger)+\frac{\sqrt{2}\xi_3}{3}(a_1^\dagger+a_2^\dagger-2a_3^\dagger)+S^\dagger\right]|000\rangle\label{Fan3},\nonumber\\
 A&\equiv& -\frac{p^2}{6} -\frac 13(\xi_2^2 +\xi_3^2-\xi_2\xi_3),\nonumber\\
 S&\equiv&\frac 23 \sum^3_{i<j=1}a_ia_j-\frac 16 \sum_{i=1}^3a_i^2.
\end{eqnarray}

\end{appendix}

\pagebreak

\pagebreak
\section*{Figures}

\begin{thebibliography}{10}
\bibitem{EPR1935}
A.~Einstein, B.~Podolsky, and N.~Rosen.
\newblock Can quantum-mechanical description of physical reality be considered
  complete.
\newblock {\em Phys. Rev.}, 47:777, 1935.

\bibitem{Bell1964}
J.S. Bell.
\newblock On the einstein podolsky rosen paradox.
\newblock {\em Physics}, 1:195, 1964.

\bibitem{Bohm1951}
D.~Bohm.
\newblock {\em Quantum Theory}.
\newblock Constable, London, 1915.

\bibitem{NielsenandChuang1968}
M.A. Nielsen and I.L. Chuang.
\newblock {\em Quantum computation and quantum information}.
\newblock Cambridge University Press, Cambridge; New York, 1968.

\bibitem{WallsandMilburn1994}
D.F. Walls and G.J. Milburn.
\newblock {\em Quantum Optics}.
\newblock Springer-Verlag, Berlin, 1994.

\bibitem{Clauseretal1969}
J.F. Clauser, M.A. Horne, A.~Shimony, and R.A. Holt.
\newblock Proposed experiment to test local hidden-variable theories.
\newblock {\em Phys. Rev. Lett.}, 23(15):880, 1969.

\bibitem{Reid1989}
M.D. Reid.
\newblock Demonstration of the {E}instein-{P}odolsky-{R}osen paradox using
  {N}ondegenerate {P}arametric {A}mplification.
\newblock {\em Physical Review}, 40(2):913, 1989.

\bibitem{Ouetal1988}
Z.Y. Ou and L.~Mandel.
\newblock Violation of {B}ell's inequality and classical probability in a
  two-photon correlation experiment.
\newblock {\em Phys. Rev. Lett.}, 61(1):50--53, Jul 1988.

\bibitem{Ouetal1992}
Z.Y. Ou, S.F. Pereira, H.J. Kimble, and K.C. Peng.
\newblock Realization of the {E}instein-{P}odolsky-{R}osen paradox for
  continuous variables.
\newblock {\em Phys. Rev. Lett.}, 68(25):3663--3666, Jun 1992.

\bibitem{BanaszekandWodkiewicz1998}
K.~Banaszek and K.~Wodkiewicz.
\newblock Nonlocality of the {E}instein-{P}odolsky-{R}osen state in the
  {W}igner representation.
\newblock {\em Phys. Rev. A}, 58:4345--4347, 1998.

\bibitem{BanaszekandWodkiewicz1999}
K.~Banaszek and K.~Wodkiewicz.
\newblock Nonlocality of the {E}instein-{P}odolsky-{R}osen state in the phase
  space.
\newblock {\em acta physica slovaca}, 49:491, 1999.

\bibitem{FanandKlauder1994}
H-Y. Fan and J.R. Klauder.
\newblock Eigenstates of two particles' relative position and total momentum.
\newblock {\em Physical Review A}, 49(2):704--707, 1994.

\bibitem{vanLoockandBraunstein2001}
P.~van Loock and S.L. Braunstein.
\newblock Greenberger-{H}orne-{Z}eilinger nonlocality in phase space.
\newblock {\em Phys. Rev. A}, 63(2), 2001.

\bibitem{Chen2002}
Z-B. Chen and Y-D. Zhang.
\newblock Greenberger-{H}orne-{Z}eilinger nonlocality for continuous-variable
  systems.
\newblock {\em Phys. Rev. A}, 65(4):044102, Apr 2002.

\bibitem{Kuzmichetal2000}
A.~Kuzmich, I.~A. Walmsley, and L.~Mandel.
\newblock Violation of {B}ell's inequality by a generalized
  {E}instein-{P}odolsky-{R}osen state using homodyne detection.
\newblock {\em Phys. Rev. Lett.}, 85(7):1349--1353, Aug 2000.

\bibitem{Mermin1990}
N.D. Mermin.
\newblock Extreme quantum entanglement in a superposition of macroscopically
  distinct states.
\newblock {\em Phys. Rev. Lett.}, 65(15):1838--1840, Oct 1990.

\bibitem{Trifonov1998}
D.A. Trifonov.
\newblock On the squeezed states for $n$ observables.
\newblock {\em Physica Scripta}, 58(3):246--255, 1998.

\bibitem{FanandZhang1998}
H-Y. Fan and Y. Zhang.
\newblock Common eigenkets of three-particle compatible observables.
\newblock {\em Physical Review A}, 57(5):3225--3228,
  1998.

\bibitem{FanandLiu2007}
H-Y. Fan and S-G. Liu.
\newblock New approach for finding multipartite entangled state representations
  via the iwop technique.
\newblock {\em International Journal of Modern Physics A}, 22(24):4481--4494,
  2007.

\bibitem{Yurkeetal1986}
B.~Yurke, S.L. McCall, and J.R. Klauder.
\newblock {SU}(2) and {SU}(1,1) interferometers.
\newblock {\em Physical Review A}, 33(6):4033--4054, 1986.

\bibitem{AdessoandIlluminatiREVIEW2007}
G. Adesso and F. Illuminati.
\newblock Entanglement in continuous variable systems: {R}ecent advances and current perspectives.
\newblock {\em Journal of Physics A: Mathematical and Theoretical}, 40:7821--7880, 2007.

\bibitem{Wigner1932}
E.P. Wigner.
\newblock On the quantum correction for thermodynamic equilibrium.
\newblock {\em Phys. Rev.}, 40:749, 1932.

\bibitem{Hilleryetal1984}
M.~Hillery, R.F. O'Connell, M.O. Scully, and E.P. Wigner.
\newblock Distribution functions in physics: Fundamentals.
\newblock {\em Physics Reports (Review Section of Physics Letters)},
  106(3):121--167, 1984.

\bibitem{Moyal1949}
J.E. Moyal.
\newblock Quantum mechanics as a statistical theory.
\newblock {\em Proceedings of the Cambridge Philosophical Society}, 45(1):99.

\bibitem{Royer1977}
A.~{Royer}.
\newblock Wigner function as the expectation value of a parity operator.
\newblock {\em Phys. Rev. A}, 15:449--450, February 1977.

\bibitem{Fan2006}
H-Y. Fan and W-Q. Wang.
\newblock Coherent-{E}ntangled {S}tate in {T}hree-{M}ode and {I}ts
  {A}pplications.
\newblock {\em Communications in Theoretical Physics}, 46(6):975--982, 2006.

\bibitem{KlauderandSkagerstam1985}
J.R. Klauder and B-S. Skagerstam.
\newblock {\em Coherent States: applications in physics and mathematical
  physics}.
\newblock World Scientific, Singapore, 1985.

\bibitem{AdessoandIlluminati2005}
G. Adesso and F. Illuminati.
\newblock Equivalence between entanglement and optimal fidelity of CV teleportation.
\newblock {\em Phys. Rev. Lett.}, 95:150503, 2005.

\bibitem{FerraroandParis2005}
A. Ferraro and M.G.A. Paris.
\newblock Nonlocality of two-and three-mode continuous variable systems.
\newblock {\em Journal of Optics B: Quantum and Semiclassical Optics}, 7:174-182, 2005.

\bibitem{vanLoockandBraunstein2000}
P.~van Loock and S.L. Braunstein.
\newblock Multipartite entanglement for continuous variables: A quantum
  teleportation network.
\newblock {\em Phys. Rev. Lett.}, 84:3482--3485, 2000.

\bibitem{Truax1985}
D.R. Truax.
\newblock Baker-{C}ampbell-{H}ausdorf relations and unitarity of {SU}(2) and
  {SU}(1,1) squeeze operators.
\newblock {\em Physical Review D}, 31(8):1988--1991, 1985.

\bibitem{FisherNietoandSandberg1984}
R.A. Fisher, M.M. Nieto and V.D. Sandberg.
\newblock  Impossibility of naively generalizing squeezed coherent states.
\newblock {\em Physical Review D}, 29(6):1107--1110, 1984.

\bibitem{Hamilton2000}
M.W. Hamilton.
\newblock Phase shifts in multilayer dielectric beam splitters.
\newblock {\em American Journal of Physics}, 68(2):186--191, 2000.

\bibitem{Berezin1966}
F.A. Berezin.
\newblock {\em The Method of Second Quantisation}.
\newblock Academic, New York, 1966.

\bibitem{Fan1990}
H-Y. Fan.
\newblock Normally ordering some multimode exponential operators by virtue of
  the iwop technique.
\newblock {\em Journal of Physics A}, 23:1833--1839, 1990.

\bibitem{deGosson2004}
M.~{de Gosson}.
\newblock On the goodness of "quantum blobs" in phase space quantization.
\newblock {\em ArXiv Quantum Physics e-prints: quant-ph/0407129}, July 2004.

\end{thebibliography}
\end{document}